\documentclass[aip,graphicx]{revtex4-1}
\usepackage{graphicx} % For including images
\usepackage{xcolor}
\usepackage{float} % For the [H] option
\draft % marks overfull lines with a black rule on the right

\begin{document}

%\preprint{}
\title{ 0.7\,MW Yb:YAG pumped degenerate optical parametric oscillator at 2.06\,$\mu$m}


\author{Anni Li}
\affiliation{Max Planck Institute for the Science of Light, Staudstrasse 2, Erlangen, 91058, Germany.}%
\affiliation{Friedrich-Alexander-Universit{\"a}t Erlangen-N{\"u}rnberg, Staudstrasse 7, 91058 Erlangen, Germany.}%

\author{Mehran Bahri}%
\affiliation{Max Planck Institute for the Science of Light, Staudstrasse 2, Erlangen, 91058, Germany.}%
\affiliation{Friedrich-Alexander-Universit{\"a}t Erlangen-N{\"u}rnberg, Staudstrasse 7, 91058 Erlangen, Germany.}%

\author{Robert M. Gray}%
\affiliation{Department of Electrical Engineering, California Institute of Technology, Pasadena, CA, 91125, USA}%

\author{Seowon Choi}%
\affiliation{Max Planck Institute for the Science of Light, Staudstrasse 2, Erlangen, 91058, Germany.}%
\affiliation{Friedrich-Alexander-Universit{\"a}t Erlangen-N{\"u}rnberg, Staudstrasse 7, 91058 Erlangen, Germany.}%

\author{Sajjad Hoseinkhani}%
\affiliation{Max Planck Institute for the Science of Light, Staudstrasse 2, Erlangen, 91058, Germany.}%

\author{Anchit Srivastava}%
\affiliation{Max Planck Institute for the Science of Light, Staudstrasse 2, Erlangen, 91058, Germany.}%
\affiliation{Friedrich-Alexander-Universit{\"a}t Erlangen-N{\"u}rnberg, Staudstrasse 7, 91058 Erlangen, Germany.}%

\author{Alireza Marandi}%
\affiliation{Department of Electrical Engineering, California Institute of Technology, Pasadena, CA, 91125, USA}%

\author{Hanieh Fattahi}%
 \email{hanieh.fattahi@mpl.mpg.de.}
\affiliation{Max Planck Institute for the Science of Light, Staudstrasse 2, Erlangen, 91058, Germany.}%
\affiliation{Friedrich-Alexander-Universit{\"a}t Erlangen-N{\"u}rnberg, Staudstrasse 7, 91058 Erlangen, Germany.}%

\begin{abstract}
Frequency comb and field-resolved broadband absorption spectroscopy are promising techniques for rapid, precise, and sensitive detection of short-lived atmospheric pollutants on-site. Enhancing detection sensitivity in absorption spectroscopy hinges on bright sources that cover molecular resonances and fast signal modulation techniques to implement lock-in detection schemes efficiently. Yb:YAG thin-disk lasers, combined with optical parametric oscillators (OPO), present a compelling solution to fulfill these requirements. In this work, we report on a bright OPO pumped by a Yb:YAG thin-disk Kerr-lens mode-locked oscillator delivering 2.8\,W, 114\,fs pulses at 2.06\,$\mu$m with an averaged energy of 90\,nJ. The OPO cavity operates at 30.9\ MHz pulse repetition rates, the second harmonic of the pump cavity, allowing for broadband, efficient, and dispersion-free modulation of the OPO output pulses at 15.45\,MHz rate. With 13\% optical-to-optical conversion efficiency and a high-frequency intra-cavity modulation, this scalable scheme holds promise to advance the detection sensitivity and frontiers of field-resolved spectroscopic techniques.

\end{abstract}

\maketitle

\section{Introduction}

Highly sensitive and precise monitoring of short-lived climate pollutants in the atmosphere is critical for advanced studies on the carbon cycle, greenhouse gas balances, and disruptions. For example, methane is a key player among greenhouse gases for its substantial role in amplifying global warming and altering climatic patterns \cite{nisbet2014methane,caulton2014toward}. Although its presence in the atmosphere is considerably less than that of carbon dioxide —about 1.8\,parts per million compared to carbon dioxide's roughly 400 ppm—methane's effect on global warming is markedly more intense, with a warming potential of 25 times greater than carbon dioxide, accounting for about 15\% of future global warming scenarios. Additionally, methane is a critical component in the intricate feedback loops of atmospheric chemistry, acting as a signal of the Earth's atmospheric dynamics on a large scale \cite{rodhe1990comparison, isaksen2011strong}.

Time-domain broadband absorption spectroscopy offers great potential for on-site, highly precise, and sensitive short-lived pollutant detection. It benefits from fast, self-calibrating assessments that do not require sample preparation, facilitating real-time atmospheric monitoring \cite{mappe2013quantum, berman2012greenhouse, grossel2007optimization, liu2015highly}. Atmospheric pollutants cover a broad fundamental vibration–rotation transition in the mid-infrared and fingerprint region. Moreover, their overtone and combination bands at short-wavelength infrared (SWIR) offer a sensitive detection window due to the lower absorption cross-section of water in this spectral range. For instance, methane exhibit distinct absorption at 1.6\,$\mu$m, and 2.2\,$\mu$m, where the water response is negligible \cite{douglass2011construction}. 

Based on the Beer-Lambert law, the detection signal-to-noise ratio in absorption spectroscopy is proportional to the brightness of the illumination source. Therefore, the detection sensitivity can be significantly enhanced by employing high-power sources operating at megahertz repetition rates. Moreover, when coupled with broadband, high-frequency modulation lock-in techniques, such frontends facilitate a further boost in detection signal-to-noise and sensitivity. Recent progress in frequency comb spectroscopy \cite{doi:10.1126/sciadv.abe9765, Malarich:23, Furst:24, zhong2024photon} and field-resolved spectroscopy \cite{herbst2022recent,srivastava2023near, scheffter2024compressed} from the SWIR range up to ultraviolet spectral range have demonstrated an unparalleled ability to identify and classify molecular responses with exceptional sensitivity and accuracy. The detection sensitivity in both techniques benefits directly from the availability of bright sources with intrinsic broadband modulation. 

Femtosecond optical parametric oscillators (OPO) synchronously pumped by Ti:Sa oscillators or fiber lasers have been a crucial tool for expanding wavelength coverage of frequency combs, as the generated sub-harmonics are locked in frequency and phase to the input pump pulses \cite{cizmeciyan2009kerr, sun2007composite, wong2010self, adler2009phase, sorokin2007sensitive, vasilyev2019middle}. Mainly when operated at degeneracy, the signal and idler pulses with the same polarization become indistinguishable, exhibiting self-phase-locked behavior through mutual injection, providing a single coherent broadband output centered at degeneracy \cite{leindecker2012octave, marandi2012coherence}. On the other hand, diode-pumped Yb:YAG thin-disk oscillators are capable of delivering sub-100\,fs pulses with tens of microjoule energy and hundreds of watts average power \cite{AusderAu:00, Saraceno:12, Fattahi:14, Goncharov:23, barbiero2021efficient, Fischer:21}. When used to pump OPOs, these advanced sources could potentially enable SWIR femtosecond sources to achieve peak powers in the MW range and average powers in the tens of watts, operating at megahertz repetition rates. The spectral coverage of such powerful OPOs can be extended into fingerprint regions or even terahertz frequencies through down-conversion processes. Furthermore, this scheme allows for high-bandwidth, dispersion-free modulation of output pulses at megahertz rates by precisely controlling the cavity length mismatch between the OPO and its pump laser \cite{Esteban-Martin:09, jiang2002harmonic}. This capability holds significant promise for enhancing detection sensitivity in spectroscopic applications, particularly in real-time atmospheric monitoring.

In this work, we present a degenerate OPO pumped by a Kerr-lens mode-locked Yb:YAG thin-disk oscillator delivering 114\,fs, 2.8\,W pulses at 30.9\,MHz repetition rates with 90\,nJ averaged pulse energy. By operating the OPO cavity at second-harmonics of the repetition rate of the pump cavity, broadband modulation of the output pulses is achieved. 

\section{Experimental results}
 \begin{figure}[t]
        \centering
        \includegraphics[width=1\linewidth]{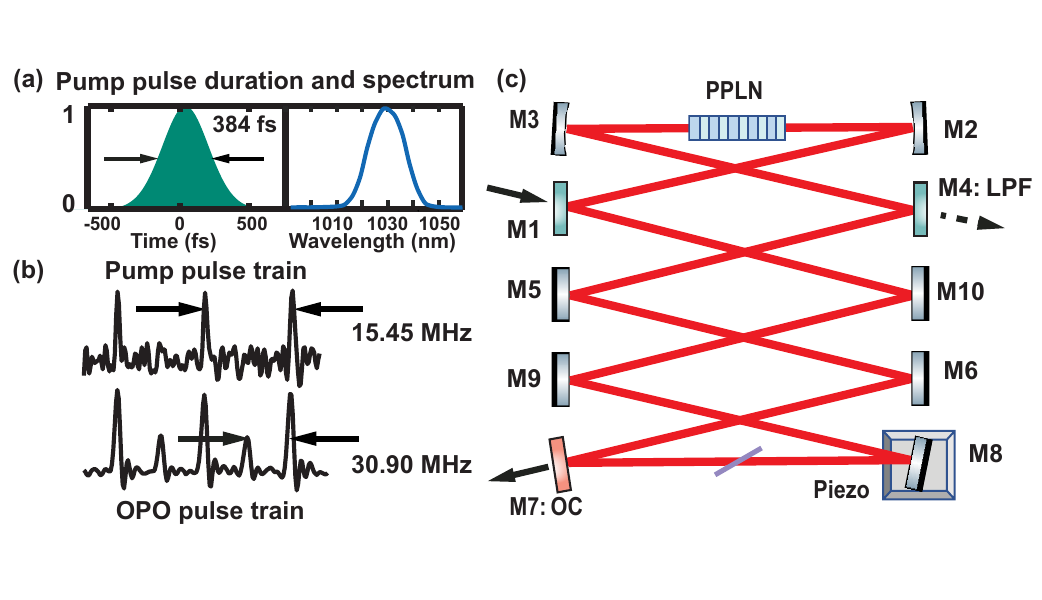}
        \caption{ (a) The temporal profile of the Yb:YAG thin-disk oscillator retrieved from a second-harmonic, frequency-resolved optical gating and its spectrum. (b) The degenerate OPO operates at the second harmonic of the pump laser repetition rate at 30.9\,MHz, while the output pulses are modulated at the pump repetition rate. (c) Schematic of the OPO cavity. M1, M4: in and output coupler of the pump laser; M2 and M3: curved mirror; M5, M6, M8, M9, and M10: folding silver-coated mirrors; M8: folding mirror clamped on a piezo stage; M7: cavity output coupler; LPF: low pass filter.}
        \label{fig:opo}
        \end{figure}
The degenerate femtosecond OPO is pumped collinearly by a Kerr-lens mode-locked Yb:YAG oscillator, delivering 25\,W, 15.45\,MHz, 384\,fs pulses at full-width at half maximum (FWHM), at 1030\,nm (see Fig. \ref{fig:opo}-a)\cite{Fattahi:16}. The pump mode is matched to the OPO cavity by a Kepler telescope with a ratio of 170\%, resulting in a focus size of 74\,$\mu$m at FWHM. This configuration corresponds to 61\,GW/$cm^2$ pump peak intensity at the focus. 

The OPO is designed for a ring cavity at the second harmonic of the repetition rate of the pump with a total length of 10\,m (Fig. \ref{fig:opo}-b). The ring cavity allows for easy in and out-coupling of the pump beam. Two concave sliver mirrors (M2 and M3 in Fig.\ref{fig:opo}-c) with a radius curvature of 700\,mm, are used to focus the cavity mode to the beam waist of 83\,$\mu$m (FWHM). By having the cavity mode 1.1 times larger than the pump beam, the efficient energy transfer from the pump to the signal is ensured. Two dielectric plane mirrors with low group delay dispersion are used for in-coupling and out-coupling of the pump beam (M1 and M4 in Fig.\ref{fig:opo}-c). A dielectric mirror with 30\% transmission is used to couple out the generated sub-harmonic pulses. Several highly reflecting silver mirrors with a protective coating fold the cavity. A 1-mm thick 5\% magnesium-doped periodically poled $LiNbO_3$ (PPLN) crystal at type-0 phase matching is used for degenerate down-conversion of the pump pulses to 2.06\,$\mu$m. The PPLN has a period of 30.8\,$\mu$m and is placed in an oven at the temperature of 116\,$^{o}C$. The end-faces of the crystal are coated by broadband anti-reflection coating with less than 0.5\% reflectivity. The PPLN is placed behind the focus of the pump beam to avoid optical damage in the crystal.

Due to the large cavity length and environmental effects, active stabilization of the cavity is required, which is achieved through Pound–Drever–Hall modulation\cite{luda2019compact} using a red-pitaya-based proportional-integral-derivative (PID) lock-in system. The feedback signal is measured at the OPO output using a photodiode. The lock-in system comprises the microcontroller board (STEMlab 125-14 board), voltage controller, main piezoelectric transducer (PZT), main piezo controller, and a supplementary PZT chip. Both PZT are affixed and bonded to one of the folding mirrors in the OPO cavity (mirror M8 in Fig.\ref{fig:opo}-c). The primary purpose of the first PZT is to coarsely scan the cavity length on a micrometer range at an approximate speed of $\sim$ 0.2 $\mu$m/s via the Red Pitaya interface.

 \begin{figure}[t]
        \centering
       \includegraphics[width=1\linewidth]{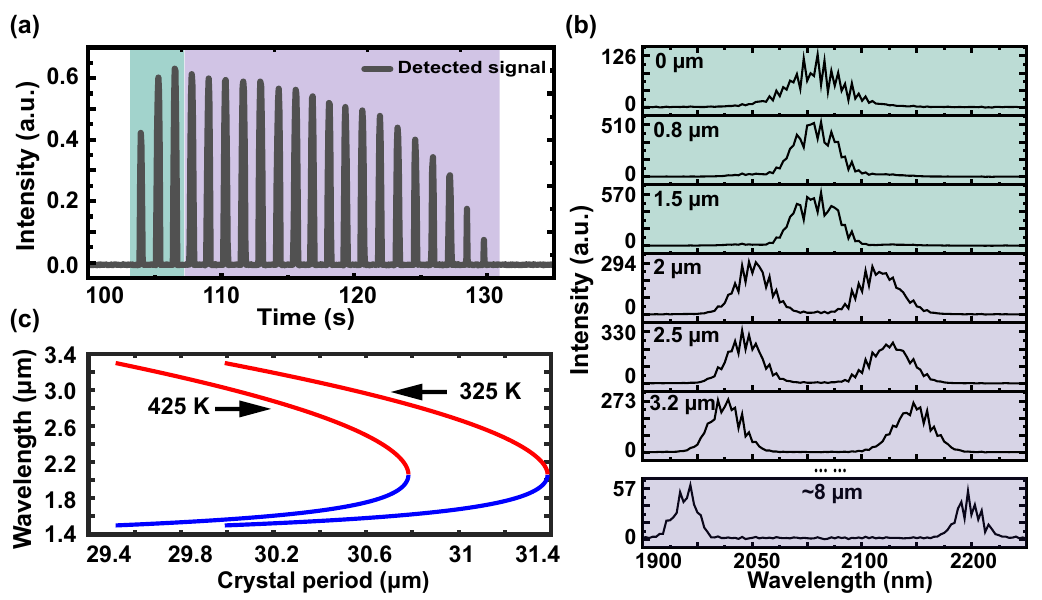}
        \caption{ (a) The measured excited cavity modes of OPO versus the cavity length scan. The time axis corresponds to the ramping signal of the PZT. The green-shaded area shows the degenerate modes, while the violet-shaded area shows the non-degenerate modes.  (b) The corresponding spectra of the OPO output of various excited modes in the cavity via cavity length tunning. The green-shaded area shows the degenerate modes, while the violet-shaded area shows the non-degenerate modes. (c) The simulated output tunability of the cavity by temperature tuning. At a crystal temperature of 425\,K, the degeneracy at 2.06\,$\mu$m occurs with a period pitch of 30.8\,$\mu$m. At a crystal temperature of 325\,K, the degeneracy at 2.06\,$\mu$m occurs with a period pitch of 31.4\,$\mu$m. The red curves show the idlers' wavelengths, while the blue curves indicate the signals' wavelengths.}
        \label{fig:detune}
       \end{figure}

Fig. \ref{fig:detune}-a shows the excited cavity modes versus the scanning range of the second PZT. The first three modes indicated by the green-shaded area are degenerate, while the subsequent non-degenerate modes are shown in the violet-shaded area. The proportional and integral functions of the PID controller are used to generate the feedback signal, with the derivative function deactivated to prevent the possible amplification of undesirable noise. Fig. \ref{fig:detune}-b presents the spectra of various excited modes within the cavity as its length is tuned. Adjusting the cavity length allows the output pulse spectrum to be tuned from 1900\,nm to 2200\,nm, constrained by the reflectivity of the cavity's internal optics. The wavelength tunability of the cavity can also be achieved via different phase-matchings in the nonlinear crystal. Fig. \ref{fig:detune}-c shows the output tunability of the cavity by phase matching of the PPLN at two periods of 30.8\,$\mu$m and 31.4\,$\mu$m, and various temperatures.
  \begin{figure}[t]
        \centering
        \includegraphics[width=0.95\linewidth]{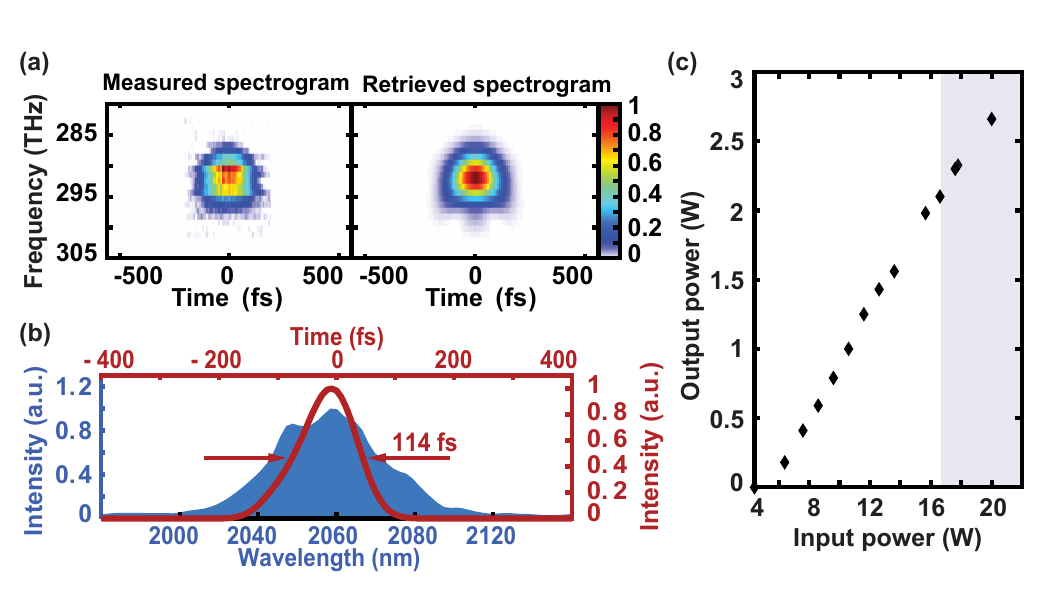}
        \caption{(a) The measured and reconstructed spectrograms of the OPO pulses characterized by the SHG-FROG. (b) The retrieved temporal profile and the spectrum of the OPO output at degeneracy. (c) The output power scaling of the OPO versus input power corresponds to 13\% optical-to-optical efficiency.}
        \label{fig:f3}
        \end{figure}

At the degeneracy, the cavity operates at negative dispersion with an estimated total group velocity dispersion of -88\,fs$^{2}$. For intra-cavity compression of the degenerate OPO pulses, a 0.9\,mm-thick Borosilicate plate is inserted into the cavity at a Brewster angle. The output pulses of the OPO are characterized by using second-harmonic generation frequency-resolved optical gating (SHG-FROG) comprising a 20\,$\mu$m-thick BBO crystal. Fig. \ref{fig:f3}-a and Fig. \ref{fig:f3}-b shows the measured and retrieved spectrograms along the temporal profile of the OPO pulses corresponding to 114\,fs at FWHM when the cavity runs at degeneracy. The spectral bandwidth of the OPO pulse supports 110\,fs Fourier transform-limited pulses. Fig. \ref{fig:f3}-c shows the output power of the OPO versus the input pump power. The output power of the OPO is measured after two long-pass filters to exclude the contribution of the other nonlinear parasitic effects in the crystal, which are co-propagating with the OPO pulses. With a 30\% output coupler, the OPO has 13\% optical-to-optical efficiency and a lasing threshold at 4\,W pump power. The purple area in the curve indicates that the cavity is approaching saturation. %with the OPO pulse-to-pulse stability of RMS as 3\%, compared with the pump pulse-to-pulse stability as 1\%. %for the first round trip

The cavity length of the OPO is half that of the pump laser, resulting in the generation of an OPO pulse upon the first incident pump pulse on the nonlinear crystal. This initial OPO pulse travels within the cavity until the arrival of the second pump pulse. During each round trip within the cavity, the OPO output pulses are coupled out through the output coupler. As the cavity operates at the second harmonics of the pump pulses, the OPO pulse meets the pump pulse and obtains parametric gain after traveling two round trips in the OPO cavity. Therefore, the output pulse train of the OPO has a repetition rate of 30.9\,MHz, which is determined by the cavity length of the OPO. The time interval between every two consecutive pulses in the train is 324 $\mu$s. At this regime, the cavity output is modulated at 15.45\,MHz with a broadband modulation depth of 41\% (Fig. \ref{fig:opo}-b). The modulation depth between the two adjunct pulses can be tuned by varying the cavity loss in the OPO cavity. For small cavity losses, the energy of the succeeding pulses will be only a few percent less than that of the previous one. The inherent capability for adjustable broadband modulation makes such cavities an ideal frontend for highly sensitive spectroscopic applications.

\section{Conclusion}

 \begin{figure}[t]
        \centering
        \includegraphics[width=0.8\linewidth]{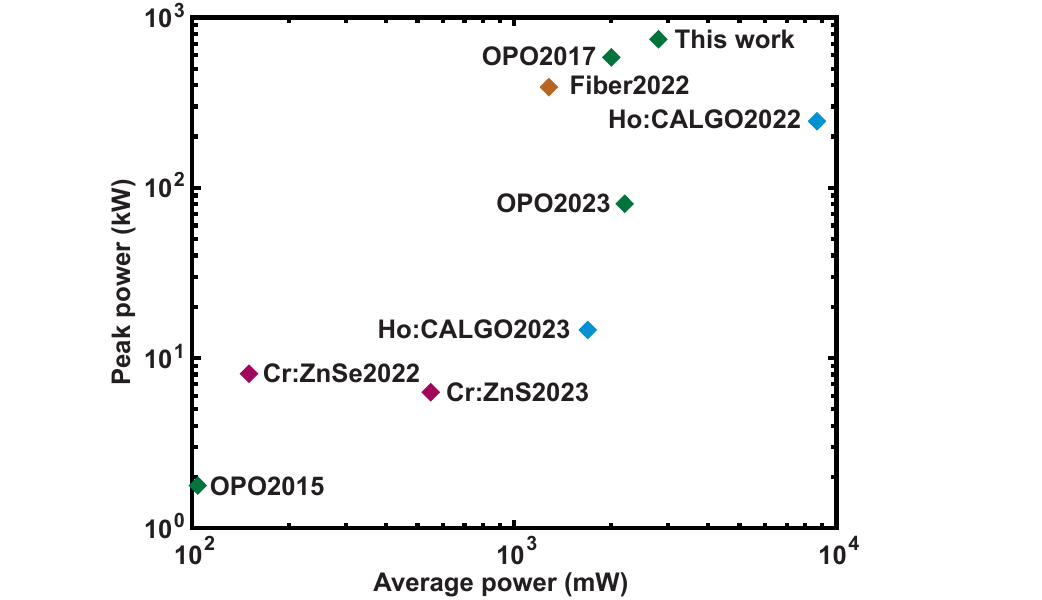}
        \caption {The state-of-the-art megahertz sources at overtone and combination band spectral range \cite{Marandi:15,WANG2022108300,Petersen:17,10015750,hoghooghi20221,Yang_2023,yao2023ghz,yao20228,shang2023short}.}
        \label{fig:methane}
        \end{figure}

The availability of broadband, bright sources from ultraviolet to terahertz is crucial for sensitive, precise, and live monitoring of short-lived climate pollutants in the atmosphere \cite{kwasny2023optical}. The fundamental resonances of molecules have a unique fingerprint in the mid-infrared spectral range, allowing for precise spectroscopic selectivity. Overtone and combination bands of molecules at SWIR provide similar information with the advantage of lower water absorption cross-section. Among various down-conversion schemes aimed at generating broadband pulses for absorption spectroscopy across overtone, fundamental, and rotational resonances, OPO stands out for its higher conversion efficiency \cite{Reiger:24, Fattahi:16, fattahi2016sub, fattahi2013efficient, alismail2017carrier}. Yb:YAG thin-disk oscillators offer great potential to pump the OPOs and to elevate the brightness, repetition rates, and peak power of the femtosecond pulses in these spectral ranges. Moreover, enhancing detection sensitivity in absorption spectroscopy requires implementing lock-in systems, necessitating high-rate signal modulation. Existing techniques like acousto-optic \cite{tournois1997acousto} or mechanical modulation suffer from limitations such as insufficient bandwidth, low-frequency modulation rate, high spectral dispersion, or low throughput.  

This work presented the first Yb:YAG thin-disk pumped OPO containing a broadband, dispersion-free modulator. The OPO comprised a ring cavity and a 1\,mm-thick PPLN crystal operating at the second harmonics of the Yb:YAG cavity at 30.9\,MHz repetition rates. The frontend delivers 114\,fs, 2.8\,W pulses at 2.06\,$\mu$m with an average energy of 90\,nJ. The wavelength tunability of the cavity from 1900\,nm to 2200\,nm is demonstrated by adjusting the cavity length, overlapping with several crucial atmospheric gasses resonances in SWIR, particularly with the water-free window of Methane. Fig. \ref{fig:methane} compares the performance of the frontend with the state-of-the-art sources at MHz repetition rates at overtone and combination band spectral regions. The intrinsic broadband, dispersion-free signal modulation at 15.45\ MHz offers a great opportunity for enhancing detection sensitivity in field-resolved spectroscopy. The system holds potential for further wavelength tunability by optimizing the intracavity dispersion or employing other nonlinear crystals. At the same time, the average power and peak power of the OPO pulses can be scaled by implementing higher energy Yb:YAG oscillators at MHz repetition rates, allowing for the down-conversion of the OPO pulses to terahertz and mid-infrared spectral range \cite{Wang:23, barbiero2020broadband}. With an intrinsic modulator, this scalable scheme holds promise to pave the path for atmospheric monitoring of short-lived pollutants and novel spectroscopic techniques \cite{herbst2022recent}.

\begin{acknowledgments}
We thank Kilian Scheffter for the insightful discussion. Anni Li and Mehran Bahri acknowledge scholarship from the Max Planck School of Photonics, which is supported by the German Federal Ministry of Education and Research (BMBF), the Max Planck Society, and the Fraunhofer Society.
\end{acknowledgments}

\section*{DECLARATIONS}
\begin{itemize}
    \item This work was supported by research funding from the Max Planck Society.
    \item Conflict of interest/Competing interests: The authors do not declare any competing interests.
\end{itemize}

\section*{DATA AVAILABILITY}
The data supporting this study's findings are available from the corresponding author upon reasonable request.

\bibliography{sample}
\end{document}